\documentclass[12pt]{article}

\usepackage{sidecap}
\usepackage{booktabs}   
\usepackage{multirow}   
\usepackage{amsmath}    
\usepackage{scicite}
\usepackage{geometry}
\usepackage{rotating} 

\usepackage{subcaption}

\usepackage{graphicx}
\usepackage{xspace}
\usepackage{txfonts}
\usepackage{xargs}

\usepackage{upgreek}
\usepackage[colorlinks=true,linkcolor=blue,citecolor=blue]{hyperref}

\usepackage{gensymb}
\DeclareRobustCommand{\ion}[2]{\textup{#1\,\textsc{\lowercase{#2}}}}
\newcommand{\jwst}{\emph{JWST}\xspace}

\newcommand{\heii}{\ion{He}{ii}\xspace}
\newcommand{\hii}{\ion{H}{ii}\xspace}
\newcommand{\hi}{\ion{H}{i}\xspace}
\newcommand{\oi}{\ion{O}{i}}
\newcommand{\oii}{\ion{O}{ii}}
\newcommand{\oiii}{\ion{O}{iii}}

\newcommand{\ci}{\ion{C}{i}}
\newcommand{\cii}{\ion{C}{ii}}
\newcommand{\ciii}{\ion{C}{iii}}
\newcommand{\civ}{\ion{C}{iv}}
\newcommand{\nv}{\ion{N}{v}}
\newcommand{\niv}{\ion{N}{iv}}
\newcommand{\niii}{\ion{N}{iii}}
\newcommand{\alii}{\ion{Al}{ii}}
\newcommand{\sii}{\ion{Si}{ii}}

\newcommand{\feii}{\ion{Fe}{ii}}
\newcommand{\siv}{\ion{Si}{iv}\xspace}
\newcommand{\lya}{Ly$\alpha$\xspace}

\newcommandx{\citep}[3][1=,2=]{%
  \if\relax\detokenize{#1}\relax
  \else
    , #1 ref.\ 
  \fi
  (%
    \textit{\citen{#3}}%
    \if\relax\detokenize{#2}\relax
    \else
      , #2 ref.\ %
    \fi
  )%
}

\newcommand{\citet}[1]{\cite{#1}}

\newcommand{\myabstract}[1]{%
  \begin{quote}\bfseries #1\end{quote}%
}

\newcounter{lastnote}

\title{Chemical signatures from the first stars embedded in metal-poor gas in galaxies at cosmic dawn}

\author{
Clara L. Pollock$^{1,2}$, 
Kasper E. Heintz$^{3,1,2}$, 
Elka Rusta$^{4,5}$, 
Darach Watson$^{1,2}$, \\
Stefania Salvadori$^{4,5}$, 
Callum Witten$^{6}$,
Joris Witstok$^{1,2}$,
Ioanna Koutsouridou$^{4}$, \\
Viola Gelli$^{1,2}$, 
Pascal A. Oesch$^{6,1,2}$, 
Gabriel B. Brammer$^{1,2}$, 
Andrea Saccardi$^{7,8}$, \\
Kei Ito$^{1,3}$, 
Rashmi Gottumukkala$^{1,2}$,
Dagmar Bergholt$^{1,2}$,
Kasper R. Brooksby$^{1,2}$, \\
Chamilla Terp$^{3,1,2}$, 
Francesco Valentino$^{1,3}$
}

\date{}

\begin{document} 




\maketitle 
\begin{changemargin}{-0.2cm}{-0.2cm} 
{\footnotesize \noindent
\normalsize{$^{1}$Cosmic Dawn Center (DAWN), Denmark}\\
\normalsize{$^{2}$Niels Bohr Institute, University of Copenhagen, Blegdamsvej 17, K{\o}benhavn 2100, Denmark}\\
\normalsize{$^{3}$DTU Space, Technical University of Denmark, Elektrovej 327, DK2800 Kgs. Lyngby, Denmark}\\
\normalsize{$^{4}$Dipartimento di Fisica e Astronomia, Università degli Studi di Firenze, Largo E. Fermi 1, 50125, \\ Firenze, Italy}\\
\normalsize{$^{5}$INAF/Osservatorio Astrofisico di Arcetri, Largo E. Fermi 5, 50125, Firenze, Italy}\\
\normalsize{$^{6}$Department of Astronomy, University of Geneva, Chemin Pegasi 51, 1290 Versoix, Switzerland}\\
\normalsize{$^{7}$Université Paris-Saclay, Université Paris Cité, CEA, CNRS, AIM, 91191, Gif-sur-Yvette, France}
\normalsize{$^{8}$Centre national d’études spatiales (CNES), Paris, France}
}
\end{changemargin}

\myabstract{%
The first generation of stars formed from pristine, neutral hydrogen gas. The most massive of these exploded as supernovae within a few million years of their birth, producing the first heavier elements and leaving distinct chemical signatures of their origin in the surrounding medium. However, chemical abundance studies have so far mainly relied on emission-line measurements, which are luminosity weighted and hence biased towards the most recently formed stars. Here we analyse near-infrared, medium-resolution spectroscopy from the \jwst-SPURS program of three UV-bright galaxies at redshifts 7.8, 8.6, and 9.3, within the first 650 to 520 million years after the Big Bang. The chemical abundance patterns of the metal lines detected in absorption hint at extremely metal-poor gas, substantially lower than inferred from the emission lines tracing the central, star-forming regions. Further, they all exhibit super-solar [C/O] abundances, which is also imprinted in the averaged spectrum of a larger set of galaxies at similar redshifts. These results reveal the distinct chemical signatures of the first Population III supernovae explosions.
}


\section{Introduction}

The first stellar populations, so-called Population III (Pop\,III), formed from pristine gas accreted into the first virialised cosmic structures \citep{KlessenGlover23}. 
Evidence for these abundant reservoirs of neutral atomic \hi\ gas has previously been seen in \lya\ absorption with JWST in galaxies at high redshifts \citep{Heintz24_DLA,Umeda24,Hainline24,Pollock26_DLA}.  
Because of their pristine environment of formation, Pop\,III stars are predicted to be more massive than present-day stars \citep{Hirano15} and thus shorter lived \citep{Tumlinson00,Schaerer02}. 
The explosion of a single Pop\,III star could be enough to pollute its surroundings with enough of heavy elements produced via stellar nucleosynthesis in its core that the system is no longer considered chemically pristine, and such systems have therefore proved extremely difficult to identify. However, the exact chemical yields depend on the mass of their progenitor star and the supernova explosion energy \citep{Heger02,Heger10,Limongi18,Salvadori19} such that distinct abundance patterns from Pop\,III stars are produced and encoded in the surrounding interstellar or circumgalactic medium, providing the most direct tracer of these first massive, pristine stars.

The first searches for chemical signatures from Pop\,III stars targeted metal-poor stars in our Galaxy \citep{Frebel15,Bonifacio25}, with a distinct class of carbon-enhanced metal-poor (CEMP) stellar sources \citep{Bonifacio15} being found as the most likely direct descendants of the first generation of stars \citep{Vanni23,Koutsouridou23}. Quasar sightlines toward distant, high-redshift galaxies have also been utilised to look for similar chemical signatures imprinted in absorption from the gaseous disks \citep{Pettini08,Cooke11,Dutta14,Saccardi23,Vanni24,Sodini24}. However, this can only be achieved out to limited redshifts ($z\lesssim 5$) where the gas mainly traces the outskirts of the foreground galaxies and typically with impact parameters at unknown radii to the central stellar population. More recently, with the advent of JWST, direct searches for Pop\,III signatures have been carried out in distant galaxies, with some cases showing C/O abundance ratios in excess to what can be produced from typical pre-enriched (Pop\,II) supernovae \citep{Cameron23,DEugenio24_carbon,Nakajima25}. These are relatively easy to measure from the detections of prominent ionised nebular C and O emission line features. However, this method suffers from biases when converting the emission-line strengths to abundance measurements due to uncertain temperatures, densities, and ionisation strength in the star-forming \hii regions producing the lines. 

In this work, we present a detailed characterisation of the absorption lines from low-ion metal species detected in the medium-resolution spectra with JWST of three UV-bright galaxies at redshifts $z\approx 7.8-9.3$ (Sect.~\ref{sec:observations}). This provides a direct, unbiased measure of the metal abundances in the absorbing gas (Sect.~\ref{sec:abslines}). The chemical patterns of the three galaxies are compared to local CEMP stars and high-redshift quasar absorbers, showing very metal-poor gas with super-solar C/O abundances (Sect.~\ref{sec:results}). Combined with novel galaxy formation models, we demonstrate that the derived abundance patterns are consistent with the expected chemical yields from the first generation of supernovae from massive Pop\,III progenitors (Sect.~\ref{sec:discussion}).

\section{Observations}\label{sec:observations}

We analyse three galaxies at redshifts $z\approx 7.8 - 9.3$, and place them in connection with a stack of a larger sample (see Supplementary Text), all observed with the JWST Near Infrared Spectrograph (NIRSpec) grating spectra (G140M), covering the rest-frame UV ($\lambda \approx 1200-2000\,\AA$) at a spectral resolving power $\mathcal{R}\approx 1000$ \citep{Jakobsen22}. The three main galaxies are observed as part of the JWST Cycle 4 programme `SPURS' (ID: 9214, PIs: C. Mason and D. Stark). They are selected to have average signal-to-noise ratios ${\rm S/N}>5$ per resolution element for the continuum in the wavelength region covering the most prominent low-ion metal absorption lines from \lya\ to \alii\,$\lambda 1670$. Each galaxy has previously been observed with either the low-resolution Prism configuration or longer-wavelength high-resolution spectra (see Supplementary Text). They are among the brightest objects known at $z\gtrsim 8$ with absolute UV magnitudes brighter than $M_{\rm UV} = -20.5$\,mag. Their general physical properties are summarised in Table~\ref{tab:properties}. Due to their brightness and the long integrations delivered by SPURS, these sources show the first clear UV absorption line features in individual galaxies at redshifts $z\gtrsim 6.5$ \citep{Chen26,Zhu26,KeerthiVasan26,Nakane26}. 

\begin{figure}
    \centering
    \includegraphics[width=1.0\linewidth]{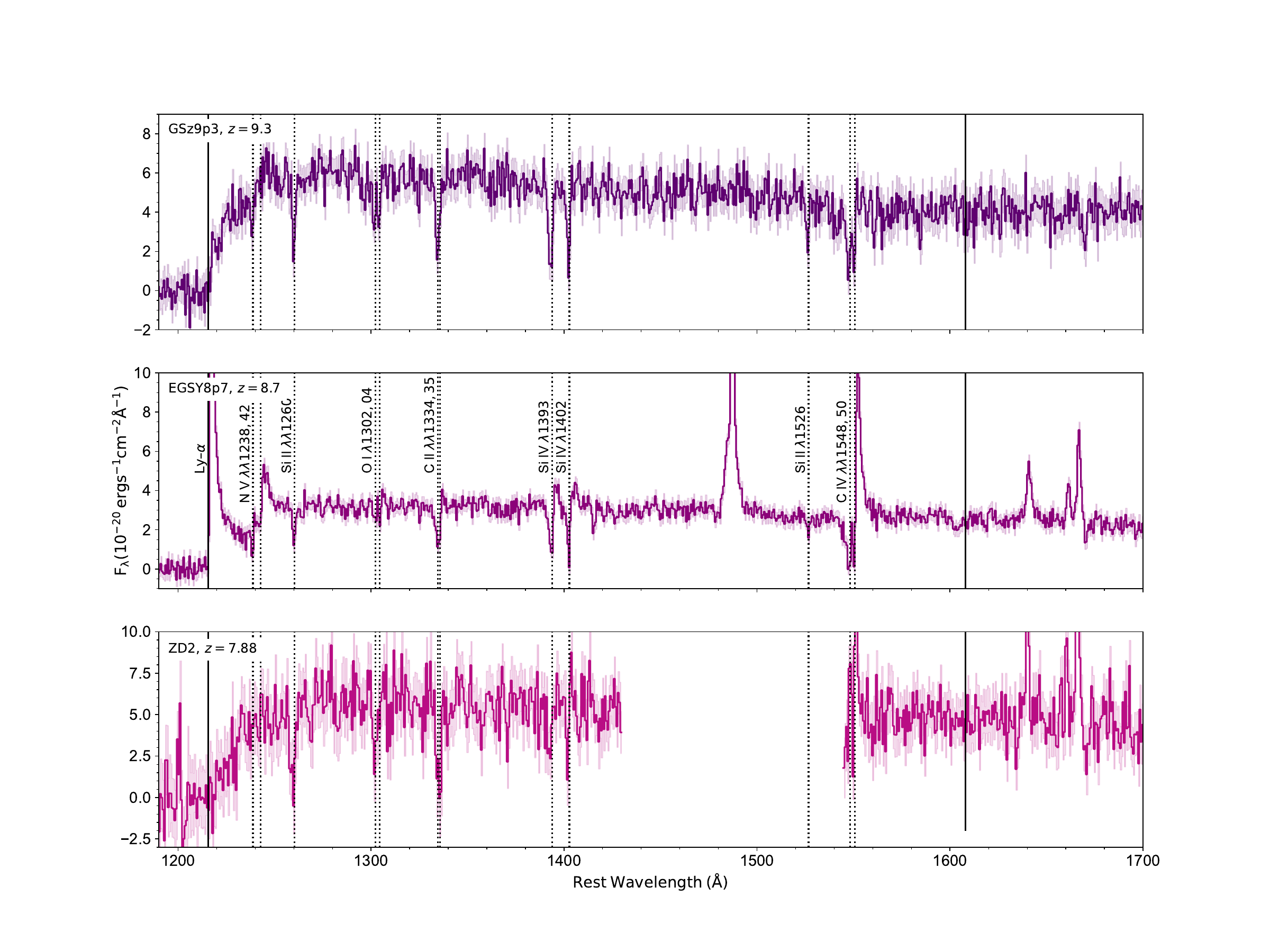}
    \caption{{\bf \jwst/NIRSpec medium-resolution rest-frame UV grating spectra of the three main galaxies.} The SPURS 1D spectroscopy of the sources studied in this work at redshifts $z=9.3113, \, 8.6817,$ and $7.8784$ are shown in purple, with detections of strong low-ion metal absorption lines highlighted.}
    \label{fig:spec_abs}
\end{figure}

The processed spectroscopic data of the three galaxies were downloaded from the DAWN JWST Archive (DJA) \texttt{v4.5} \citep{Brammer_DJA}. DJA is a compilation of all public JWST spectroscopic data, reduced in a homogeneous way optimised for high-redshift galaxies using the {\tt MSAExp} software \citep{msaexp}. The main outputs are optimally-extracted 1D spectra, with custom path loss corrections accounting for the flux outside the shutter and with extended wavelength coverage from the nominal. The exact details on the reduction and post-processing steps are detailed in dedicated papers \citep{Heintz25_PRIMAL,DeGraaff25_RUBIES,Valentino25,Pollock26_FMR}. 
The G140M grating spectra covering the rest-frame UV are shown for each galaxy in Figure~\ref{fig:spec_abs}, with the main low-ion absorption-line features highlighted. All three galaxies show a prominent \lya\ damping wing, with implied \hi\ column densities in the range $N_{\rm HI}= 10^{21}-10^{22}$\,cm$^{-2}$, in excess of the expected contribution from neutral hydrogen in the intergalactic medium (IGM, see Supplementary Text). EGSY8p7 also shows prominent \lya\ emission and several high-ionisation nebular emission lines such as \nv\,$\lambda\lambda 1238,1242$, \siv\,$\lambda\lambda 1392,1402$, \niv\,$\lambda1486$, and \civ\,$\lambda 1550$ \citep{MarquesChaves26}.

\begin{table}[]
    \centering
    \begin{tabular}{l c c c}
    \hline\hline
        & Gz9p3 & EGSY8p7 & ZD2 \\
        \hline
        R.A. (deg., J2000) & $3.617169$ & $215.035395$ & $3.604499$  \\
        Decl. (deg., J2000) & $-30.42555$ & $52.89067$ & $-30.38046$  \\
        $z_{\rm spec}$ & $9.3113$ & $8.6817$ & $7.8784$  \\
        $M_{\rm UV}$ (mag) & $-21.5 \pm 0.05$ & $-21.8 \pm 0.1$ & $-20.68 \pm 0.1$ \\
        $\beta_{\rm UV}$ & $-1.90 \pm 0.03$ & $-1.13 \pm 0.19$ & $-1.03^{+0.40}_{-0.46}$  \\
        $A_V$ (mag) & $0.10\pm 0.01$ & $0.21\pm 0.01$ & $0.13 \pm 0.02$  \\
        $T_e$ ($10^{4}$\,K) & $1.57\pm 0.31$ & $1.83\pm 0.12$ & $1.88\pm 0.19$ \\
        12+log(O/H)$_{\rm em}$ & $8.02\pm 0.25$ & $7.70\pm 0.06$ & $7.70\pm 0.10$  \\
        ${\rm [C/O]}_{\rm em}$ & $-0.62\pm 0.23$ & $-0.25\pm 0.01$ & $-0.5\pm 0.04$  \\
        $\log N$(H\,{\sc i}\,/\,cm$^{-2}$) & $21.0^{+0.3}_{-0.4}$ & $21.86^{+0.04}_{-0.05}$ & $21.73^{+0.12}_{-0.15}$\\
        ${\rm [O/H]}_{\rm abs}$ & $-2.03\pm 0.24$ & $-3.72\pm 0.14$ & $-2.64\pm 0.46$  \\
        ${\rm [C/O]}_{\rm abs}$ & $0.19\pm 0.30$ & $1.07\pm 0.16$ & $0.69\pm 0.56$  \\
        ${\rm [Si/O]}_{\rm abs}$ & $-0.26 \pm 0.29$ & $0.08 \pm 0.16$ & $0.28\pm0.55$ \\
       \hline
    \end{tabular}
    \caption{Spectroscopic properties and chemical abundances of the three SPURS galaxies.}
    \label{tab:properties}
\end{table}

\section{Absorption-line spectroscopy}\label{sec:abslines}

Probing the interstellar medium (ISM) in absorption provides a clear advantage over emission line observations, since it measures a direct abundance in the line of sight \citep{Peroux20}. This technique is also independent of redshift and is more sensitive to low-density gas and metals located beyond the regions excited by hot stars, and is unaffected by high-density collisional de-excitation that the rest-frame optical [\oiii] and [\oii] emission lines might be biased by. 
We model the absorption lines with Voigt profiles using the Python package \texttt{VoigtFit} \citep{Krogager18}. This takes as input the G140M spectra and convolves the models of the absorption-line profiles and continuum with the delivered spectral resolving power to derive column densities, $N$, as output. When available, the following set of low-ion metal features were modelled: \sii\,$\lambda1260$, \oi \,$\lambda1302$, \oi*$\lambda 1304$, \sii \,$\lambda1304$, \cii\,$\lambda1334$, \cii* $\lambda1335.7$, \sii\,$\lambda1526$, and \alii\,$\lambda 1670$. These low-ion transitions all probe the metals in the neutral gas-phase, since their ionisation potentials are below or equal to that of hydrogen (13.6 eV), which are thus effectively shielded due to the high \hi\ gas column densities. The model assumes a shared turbulent broadening `$b$'-parameter between each component, which we fixed to $b=100$\,km\,s$^{-1}$ for EGSY8p7 and ZD2 to be conservative and $b=50$\,km\,s$^{-1}$ for Gz9p3. The latter was chosen to best reproduce the line widths, as the larger broadening adopted for the two other cases were inconsistent with the observed line profiles for this particular source. The derived column densities are generally insensitive to the exact $b$-parameter assumed (within $\Delta N<0.2$\,dex) for the range $b=20-100$\,km\,s$^{-1}$
(see Supplementary Text for more details). Our choice of $b$ effectively assumes that the lines are optically thin, such that the derived abundances should mainly be treated as lower limits. In particular, if the line profiles are intrinsically saturated, have partial coverage, or the broadening is much narrower than assumed.

\begin{figure}
    \centering
    \includegraphics[width=1.0\linewidth]{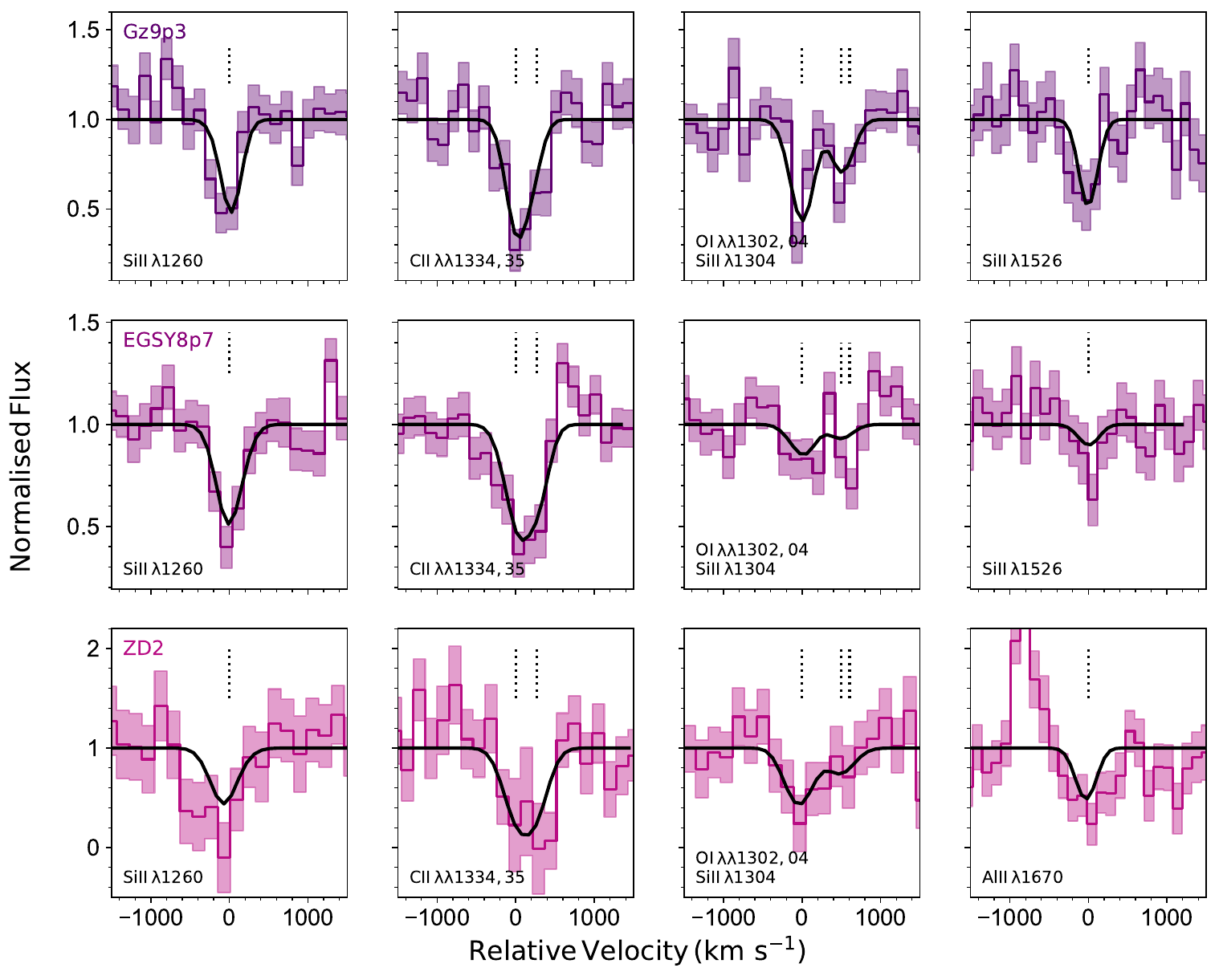}
    \caption{{\bf Normalised 1D spectra of the main absorption components.} The panels show zoom-ins on the primary low-ion absorption-line features \sii$\lambda1260$, \cii$\lambda\lambda1334,1335$, \oi$\lambda1302$, \sii$\lambda1526$, or \alii$\lambda1670$ detected in the spectra (purple) and corresponding best-fit models from \texttt{VoigtFit} (black). }
    \label{fig:zoom_abs}
\end{figure}

We show zoom-ins of each line complex in Figure~\ref{fig:zoom_abs}, along with the best-fit models from \texttt{VoigtFit}. As the lines are clearly detected we can accurately constrain the spectroscopic redshift of the absorbing system and the total column density of each species. The column densities are used to calculate specifically the [O/H] and [C/O] abundances relative to Solar \citep{Asplund21} following the typical notation: 
\begin{equation}
    {\rm [X/Y]} = \mathrm{log}\biggl(\frac{N_{\rm X}}{N_{\rm Y}}\biggl) - \mathrm{log}\biggl(\frac{N_{\rm X}}{N_{\rm Y}}\biggl)_\odot
\end{equation}
For the oxygen-to-hydrogen abundance ratio, [O/H], we use the \hi column density derived from the \lya damping wing tracing the bulk cold neutral ISM as the absorption lines (see the Supplementary Text). In all three cases, we derive super-solar [C/O]\,$>0$ abundance ratios but low metallicities at [O/H]\,$= -2.0$ to $-3.7$, as summarised in Table~\ref{tab:properties}. The low dust-to-gas ratios, $A_V/N_{\rm HI}$, of in particular EGSY8pY and ZD2 also imply low metallicities, [O/H]\,$< -1.5$, of the DLA absorbing gas \citep{Heintz23_GRB}. 
The benefit of measuring these ratios in absorption systems rather than in emission is that we require no assumptions on electron temperature or density, as the values represent the number of atoms measured along an identical line of sight. We emphasise that while the exact abundances might be underestimated, neither the potential intrinsic saturation or partial coverage will affect the [C/O] abundance measurements, since both C and O will be equally broadened and co-exist in the same neutral gas-phase. In addition, they show equivalent depletion patterns in dust \citep{Konstantopoulou24} and even have similar intrinsic transition oscillator strengths such that the relative optical depths of the integrated lines almost directly represent the actual C/O abundances. This hints at extremely metal-poor, neutral gas probed in absorption with abundance patterns that are inconsistent with those predicted from standard Pop\,II stellar yields. 

\section{Direct metallicity and C/O abundances}\label{sec:results}

\begin{figure}
    \centering
    \includegraphics[width=1.0\linewidth]{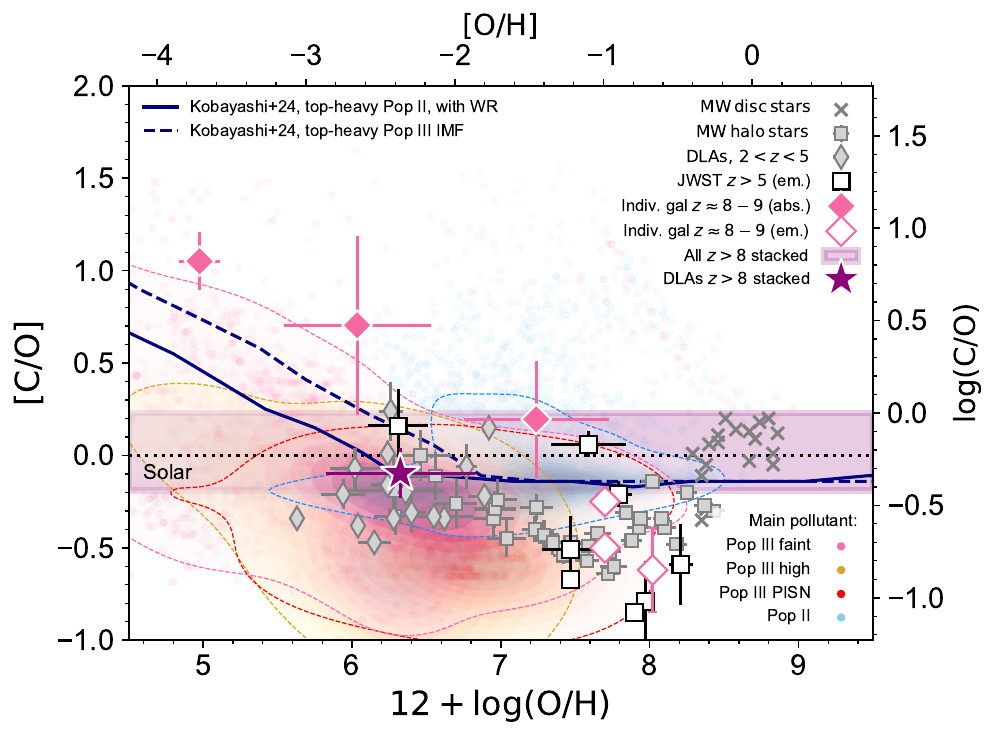}
    \caption{{\bf Evolution of the [C/O] abundance as a function of metallicity.} The absorption-derived metallicities for the three main galaxies are represented by filled pink diamonds, showing super-solar [C/O] abundances. Average values from stacked spectra $z>8$ are shown in purple, also with [C/O] abundances close to solar. We compare to literature data in grey, calculated similarly from absorption lines for quasar DLAs (diamonds) \citep{Cooke17}, Milky Way halo (squares), and disc (crosses) stars \citep{Fabbian09}. The rest-UV and optical emission-derived metallicities for the three main galaxies are shown in pink empty markers, with comparison to JWST $z>5$ literature galaxies in hollow black squares \citep{ArellanoCordova25, Jones23, Stiavelli23, Curti25, DEugenio24_carbon, Nakajima25, Hsiao25}. With the exception of GS-z12 and LAP1-B (super-solar [C/O]), the high-$z$ galaxies are all in line with the local trend. We show predicted abundances from Pop\,II and Pop\,III enriched galaxies at $z=7.5-9.5$ from the {\tt NEFERTITI} simulations as coloured scatter points and contours ($90\%$ distribution). We separate by the majority enrichment source: Pop\,III faint and CCSNe (pink), Pop\,III high-energy SNe and hypernovae (yellow), Pop\,III PISNe (red), and Pop\,II SNe (blue). 
    We compare to semi-analytical tracks from \citet{Kobayashi24} for expected yields from a top-heavy Pop\,III initial mass function (IMF, $30-120\,M_\odot$), and a top-heavy Pop\,II IMF including Wolf-Rayet stars, both predicting a carbon-enhancement at low metallicities.}
    \label{fig:CO_OH}
\end{figure}

In Figure~\ref{fig:CO_OH} we show the absorption-derived [C/O] abundance ratios for the three main galaxies as a function of their metallicities, quantified as the oxygen abundance $12+\log$(O/H). The three main galaxies exhibit super-solar [C/O] ratios of [C/O]$_{\rm abs} = 0.19\pm 0.30, 1.07\pm 0.16,$ and $0.69\pm 0.56$ for Gz9p3, EGSY8p7, and ZD2, respectively, with increasing values for lower metallicity sightlines. The spectral stack of a larger sample of galaxies observed with the same JWST/NIRSpec grating setup (Supplementary Text) also reveals that these UV-bright sources are embedded in metal-poor gas, on average ${\rm [O/H]}=-2.37\pm 0.50$, and exhibit super-solar [C/O] abundances of $0.25\pm0.01$. 
For comparison, we also show measurements from carbon-enhanced metal-poor (CEMP) stars in the Milky Way and lower-redshift damped \lya\ absorption (DLA) systems \citep{Saccardi23} in Figure~\ref{fig:CO_OH}. It is evident that the absorption-line metal abundances derived for the three main galaxies at $z\approx 8-9$ and the spectral stack are generally consistent with the [C/O]-[O/H] trend observed for Galactic CEMP stars and higher-redshift quasar absorbers. 

To compare the absorption-line abundances with the more standard high-redshift, emission-line inferences for each galaxy we calculate the oxygen abundances using the `direct' $T_e$-based method, relying on the rest-optical lines measured in the available G395M spectra and the \texttt{PyNeb} package \citep{Luridiana15}. In particular, we utilise the [\oiii] $\lambda4363$ auroral line to obtain an electron temperature $T_e$, and H$\beta$, [\oiii]\,$\lambda\lambda4959,5007$ and [\oii] $\lambda\lambda3726,3729$ fluxes to derive $\mathrm{O^{++}/H^+}$ and $\mathrm{O^+/H^+}$ contributions. We find emission-based oxygen abundances of $12+\log {\rm (O/H)} = 7.7-8.0$, corresponding to metallicities $10-20\%$ of solar. We note that strong-line diagnostics result in similar values for each galaxy, within 1$\sigma$ error. The emission line constraints are more typical to what has previously been observed at high-redshifts \citep{Heintz23,Nakajima23,Curti24}, though they are more than an order of magnitude higher than the absorption derived metallicities.
To obtain a comparable estimate of the emission-line [C/O] abundance ratio we assume the same electron temperature $T_e([\mathrm{\oiii}])$, and use rest-UV line fluxes \ciii] $\lambda1909$, \oiii] $\lambda\lambda 1660,1666$, and \civ \,$\lambda\lambda1548,1550$ (see Supplementary Text). 
This yields emission-derived [C/O] abundance ratios that are subsolar, with [C/O] $=-0.6$ to $-0.25$. We caution, however, that the emission-line carbon abundance may be overestimated if there is a hard ionisation field more effectively populating the high-ionisation states relative to oxygen. The oxygen lines could also be collisionally de-excited in dense media, as would be the case for active galactic nuclei (AGN, see the discussion in the Supplementary Text). 

The emission-derived [C/O] abundances for all three galaxies are also $1-1.5$ magnitudes lower than the absorption counterparts, though in agreement with emission-line C and O abundance measurements from other galaxies at redshifts $z>5$ \citep{ArellanoCordova25, Curti25, Jones23, Topping25, Stiavelli23}, corresponding to the predicted yields Type II core-collapse supernovae \citep{Kobayashi20}. Combined, the absorption and emission derived values for C/O and O/H reveal a two-phase medium of the target galaxies, with absorption probing more diffuse, metal-poor gas in the outskirts of the galaxies, and emission the central, more metal-rich gas. These observations are consistent with the canonical `inside-out' star formation scenario. 

\section{Implications for the nature of Pop\,III stars}\label{sec:discussion}

The observation of this metal-poor gas surrounding high-redshift galaxies is consistent with several candidate systems being observed offset from central galaxies, proposed to be metal-free Pop\,III-like companions. These include the \heii $\lambda1640$ clump \textit{Hebe} detected 20kpc from GN-z11 at $z=10.6$ with no corresponding metal lines \citep{Maiolino24_popIII, Maiolino26_hebe, Rusta26, Ubler26}, and a potential He-emitting clump offset from a nebular dominated galaxy at $z\approx5$ \citep{Reumert26}. Lensed systems with weak rest-frame optical [\oiii] lines \citep{Vanzella23, Nakajima25, Fujimoto25, Morishita25,Hsiao25} at $z\approx6.5$, and an extremely metal-poor galaxy ($\sim1.6\%$ solar) \citep{Cullen25} have also been interpreted as containing potential Pop\,III stars. However, it is difficult to categorise a purely pristine Pop\,III galaxy solely from upper limits on metal lines, as the gas metallicity would have to be substantially lower at $\lesssim 10^{-6} \, \mathrm{Z_\odot}$ \citep{Katz23, KlessenGlover23}. 

As a consequence, purely pristine star formation is extremely challenging to find. However, Pop\,III stars will imprint unique chemical signatures on the surrounding pristine gas. The exact chemical yields are determined by the initial masses of the progenitor stars and their supernova explosion energies which are still unknown \citep{Koutsouridou23,Rusta26}. Notably, low-energy core-collapse supernovae (CCSNe) from intermediate stars (i.e. $E_{SN} \lesssim 10^{51}$ erg, $M_\star \approx 10-100\, M_\odot$) are expected to expel mainly their outer C-rich layers, due to strong fallback on the remnant \citep{UmedaNomoto2003}, imprinting very high, super-solar carbon abundances. On the other hand, very massive Pop\,III stars ($M_\star \approx 140-260 M_\odot$) are predicted to explode as energetic pair instability supernovae (PISNe), leaving no remnant and imprinting a strong odd-even effect \citep{HegerWoosley2002}, with high abundances of even-Z elements such as silicon \citep{Vanni24}. 

To put our results into context, we show in Figure~\ref{fig:CO_OH} predictions from the {\tt NEFERTITI} galaxy formation model (\citep{Koutsouridou26}, see Supplementary Text for details) that trace the chemical evolution of galaxies from a fully pristine Pop\,III origin, through the increasingly enriched `hybrid phase,' to a stage dominated solely by Pop\,II stellar yields \citep{Rusta25}. This model is calibrated based on local stellar measurements and follows the star formation and chemical evolution initiated by the first stars in a self-consistent way (see Supplementary Text).
We plot galaxies from the models in the redshift range $7.5<z<9.5$ to match the regime of the observations. The chemical abundances of the metal-poor gas traced by the absorption-line measurements for the three main galaxies occupy the region of the Pop\,III-enriched phase in the [C/O]-[O/H] parameter space, though with some overlap with the Pop\,II-enrichment contours for Gz9p3 and ZD2. Searching the {\tt NEFERTITI} simulation for individual matches to the three main galaxies, we indeed find that these two can be reproduced by both Pop\,III- and Pop\,II-enriched galaxies, also taking into account their silicon abundances (Supplementary Text). On the other hand, EGSY8p7 can only be reproduced by enrichment from faint Pop\,III supernovae, revealing that the metal-poor gas surrounding this source is almost purely enriched by Pop\,III stars. This is particularly interesting in the context that this galaxy also shows emission and stellar wind signatures of very massive stars \citep{MarquesChaves26}. The comparable low [Si/O] and high [Fe/C] ratios (Supplementary Text, and Fig.~A4) further hint at the main progenitor stars having intermediate masses, $M = 10-100\,M_\odot$, and exploding as low-energy, faint supernovae with $E_{\rm SN} = 0.3-0.6\times 10^{51}$\,erg \citep{Vanni24}. These results provide the first direct measurements of the metal abundances in the gas surrounding massive, UV-bright galaxies at $z\approx 8-9$, ubiquitously representing the chemical yields expected from the first Pop\,III stellar populations.



\newpage

\bibliography{refs}

\noindent\textbf{Acknowledgements.} 
We would like to thank Jorryt Matthee and Rohan Naidu for insightful discussions during the early stages of this work. We would also like to thank Charlotte Mason and Dan Stark for constructive discussions on the metal abundances and the higher-resolution grating spectra partly analysed in this work. We are grateful to the entire SPURS team for their vision and great effort into securing the data that is analysed in this work and key to obtain the presented results. 

The DAWN \jwst\ Archive (DJA) is an initiative of the Cosmic Dawn Center (DAWN), which is funded by the Danish National Research Foundation under grant DNRF140. This work is based in part on observations made with the NASA/ESA/CSA James Webb Space Telescope. The data were obtained from the Mikulski Archive for Space Telescopes (MAST) at the Space Telescope Science Institute, which is operated by the Association of Universities for Research in Astronomy, Inc., under NASA contract NAS 5-03127 for JWST. \\

\noindent {\bf Funding.} K.E.H. acknowledges support from the Independent Research Fund Denmark (DFF) under grant 5251-00009B. The Cosmic Dawn Center (DAWN) is funded by the Danish National Research Foundation under grant DNRF140. This project received funding from the ERC Starting grant NEFERTITI H2020/804240 (PI: Salvadori). This work has received funding from the Swiss State Secretariat for Education, Research and Innovation (SERI) under contract number MB22.00072, as well as from the Swiss National Science Foundation (SNSF) through project grant 200020\_207349. A.S. acknowledges financial support from the Centre national d’études spatiales (CNES), France (ROR: \url{https://ror.org/04h1h0y33}) within the framework of the SVOM mission.  \\

\noindent {\bf Author Contributions.} K.E.H. and D.W devised the main idea for this work. C.L.P. and K.E.H. drafted the manuscript and led the analysis. E.R. and I.K. produced the presented NEFERTITI simulations output, under guidance from S.S. and V.G. C.W. helped interpret the absorption-line spectroscopy, and D.B. and K.B. reproduced the analysis. G.B. reduced the observations and extracted the spectra. K.I. and R.G. performed the spectral energy distribution modelling of the sources. J.W., P.A.O., A.S., C.T., and F.V. contributed with key insights to the interpretation and analysis of the results. \\ 

\noindent {\bf Competing interests.} The authors declare no competing interests. \\

\noindent {\bf Data and materials availability.} The JWST spectroscopic data are available at the Mikulski Archive for Space Telescopes (MAST) \url{https://mast.stsci.edu}. Partly based on data obtained as part of JWST program with ID 9214. The relevant proposal number and source IDs are provided for each source in the Supplementary Text. The reduced spectroscopic data are available through
the DJA: \url{https://s3.amazonaws.com/msaexp-nirspec/extractions/nirspec_public_v4.5.html}.


\appendix
\setcounter{figure}{0}
\renewcommand{\thefigure}{\thesection\arabic{figure}} 
\renewcommand{\figurename}{Fig.} 
\newpage
\section{Supplementary Text}\label{app:suppsection}

\subsection{Cosmology}
Throughout the paper, we assume the standard cosmology, with a flat, $\Lambda$ cold dark matter-dominated Universe. We adopt the most recent cosmological measurements from the Planck Collaboration \citep{Planck18}, with a Hubble constant $H_0 = 67.4$\,km\,s$^{-1}$\,Megaparsec$^{-1}$, matter density parameter $\Omega_{\rm m} = 0.315$, and dark energy density parameter $\Omega_\Lambda = 0.685$. 
We further assume the solar abundances from Asplund et al. \citep{Asplund21} with $\mathrm{12 + log(O/H)}_\odot = 8.69$, $\mathrm{log(C/O)_\odot} = -0.23$, and $\mathrm{log(Si/O)_\odot} = -1.18$ as the benchmark for the high-redshift metal-abundance ratios and the chemical evolutionary tracks.

\subsection{Observations and data reduction}

The main three galaxies examined in this work have been extensively studied in the literature in the past, with substantial auxiliary observations and data. The first galaxy, Gz9p3, is also called UNCOVER-3686 \citep{Bezanson24}, DHZ1 \citep{Algera25}, and SPURS-A2744-7 \citep{Chen26}. The rest-frame UV \jwst\ spectra of this source showed the first indications of neutral gas absorption features below the stellar continuum \citep{Boyett24,Zhu26}. It further shows a prominent detection of the far-infrared [\oiii]-$88\mu$m line from Atacama Large Millimetre/sub-millimetre Array (ALMA) observations \citep{Algera25}. The second galaxy, EGSY8p7, has until recently been one the most distant \lya\ emitters known \citep{Zitrin15,Mainali18,Larson22}, now confirmed by \jwst\ \citep{Tang23}. It is also known as RUBIES-8488 \citep{DeGraaff25_RUBIES} and CEERS-1019 \citep{Larson23,Zamora25}. The rest-frame UV \jwst\ spectrum of EGSY8p7 studied here, have also recently been shown to contain emission signatures of very massive stars and prominent stellar winds \citep{MarquesChaves26}. The third galaxy was first identified as ZD2 \citep{Zheng14} and belongs to the overdensity of galaxies at $z\approx 7.88$ discovered behind the Abell\,2744 lensing cluster \citep{Morishita23,Witten25}. It is also known as GLASS-100003, GLASSZ8-1 \citep{RobertsBorsani22}, and SPURS-A2744-17 \citep{KeerthiVasan26}. The \jwst\ rest-frame UV spectrum has recently been studied in separate works, highlighting also the presence of prominent metal absorption lines \citep{Zhu26, KeerthiVasan26}.

For this work, we mainly consider the \jwst\ spectra obtained with the G140M/F100LP grating configuration \citep{Jakobsen22}. This nominally covers $\lambda = 0.97 – 1.89\mu$m, corresponding to the rest-frame UV at these redshifts, at a spectral resolving power $\mathcal{R}=400 - 1500$. The three main galaxies studied in this work are observed as part of the JWST Cycle 4 `SPectroscopic Ultra-deep Reionization-era Survey' (SPURS) programme (GO-9214, PIs: Mason \& Stark). To supplement the SPURS data, we compile all the available G140M spectra on the DAWN \jwst\ archive (DJA; \citep{Brammer_DJA}) of galaxies at $z>8$ to make a high-S/N stack of similar spectra (see Sect~\ref{sec:stack} below) to test if the distinct absorption features are detected in an average stack of a representative set of galaxies at high-redshifts. There are 32 unique spectra in total, with 4 strong line emitters, 8 with prominent DLAs and significant continua detections, and a further 20 with only tentative features or continua detections. The data are from a range of \jwst\ programmes, including JADES (PIDs: 1181, 1286, 1287, 3215, 1210), SPURS (9214), GO-4750, and GO-5943, as summarised on the main DJA-spec online repository\footnote{\url{https://s3.amazonaws.com/msaexp-nirspec/extractions/nirspec_public_v4.5.html}}. 

The processed spectroscopic data of all the galaxies studied here were downloaded from the DJA \texttt{v4.5} \citep{Brammer_DJA}. DJA is a compilation of all public JWST spectroscopic data, reduced in a homogeneous way, optimised for high-redshift galaxies using the {\tt MSAExp} software \citep{msaexp}. The main outputs are optimally-extracted 1D spectra, with custom path loss corrections accounting for the flux outside the shutter and with extended wavelength coverage from the nominal, extending up to $\approx 2.6\mu$m for the G140M/F100LP grating configuration via the inclusion of the second-order light. The exact details on the reduction and post-processing steps are detailed in dedicated papers \citep{Heintz25_PRIMAL,DeGraaff25_RUBIES,Valentino25,Pollock26_FMR}. 

\subsection{Physical properties and classifications} \label{sec:physprop}

We model the full spectral energy distribution (SED) from rest-UV to rest-optical using the available G140M, G235M, and G395M medium-resolution ($\mathcal{R}\sim 1000$) spectra on DJA using the Python software `Bayesian Analysis of Galaxies for Physical Inference and Parameter EStimation' (\texttt{BAGPIPES}) \citep{Carnall18}. We jointly model the photometry and spectroscopy, assuming a non-parametric continuity star-formation history (with priors described in ref. \citet{Leja19}). We include a flexible dust curve \citep{Salim18}, modified from Calzetti et. al. \citep{Calzetti00}, which includes a UV dust bump $B$ at 2175$\mathrm{\AA}$, and a deviation $\delta$ from the default attenuation curve slope \citep{Calzetti00}. From the SED fitting we derive the visual dust extinction $A_V$, and also use the intrinsic spectral model to verify the robustness of the DLA fitting to the underlying continuum (see Section~\ref{sec:lya}).

\begin{figure}
    \centering
    \includegraphics[width=1.0\linewidth]{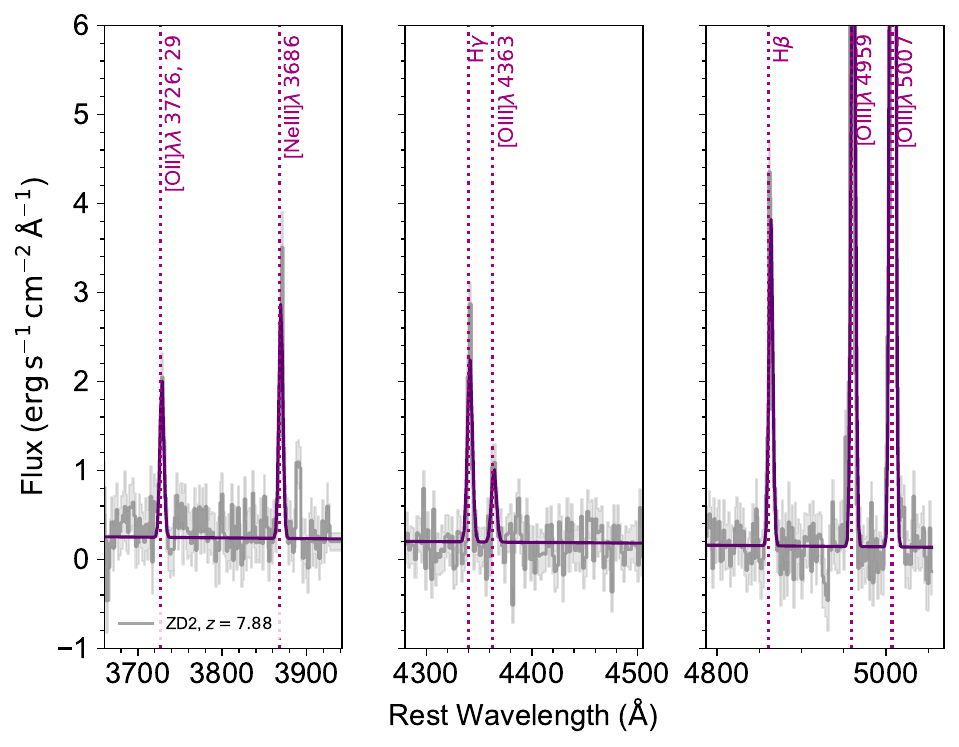}
    \caption{{\bf Example of rest-optical emission-line fitting.} The JWST/NIRSpec 1D spectrum for A2744-ZD2 at $z=7.88$ is shown in gray, with the most prominent nebular emission lines needed to calculate 12+log(O/H) using the `direct' method highlighted, including [\oii]$\lambda\lambda3726,29$, H$\gamma$, the temperature-sensitive auroral line [\oiii]$\lambda4363$, H$\beta$, and [\oiii]$\lambda\lambda4959, 5007$. }
    \label{fig:example_opt}
\end{figure}

The most prominent nebular emission line fluxes are modelled as Gaussian profiles and used to calculate the emission-derived abundances 12+log(O/H) and C/O, tracing the central star-forming regions, as highlighted in Section~\ref{sec:results}. Rest-optical lines are measured in G395M, assuming a flat underlying continuum. An example of the modelling can be seen in Figure~\ref{fig:example_opt} for ZD2 at $z=7.88$. The rest-frame UV lines are measured in the G140M spectra, with the underlying continuum assumed to be a power law (see also Sect.~\ref{sec:lya} below). To derive the emission-line $\mathrm{[C/O]}_{\rm em}$ abundance ratio we use the rest-UV emission lines \civ$\lambda\lambda1548,50$ and \ciii$\lambda1909$ relative to [\oiii]$\lambda\lambda1660,66$, but assuming an electron temperature $T_e$ derived from [\oiii]$\lambda4363$ in the rest-optical. When \civ\ is not available (overlapping with detector gap in ZD2), we assume an ionisation correction factor (ICF) calculated from the O32 ratio ([\oiii]\,$\lambda5007$/[\oii]\,$\lambda3727$) in order to account for potential higher ionisation states. 

For the three main galaxies we derived $T_e$-based emission-line oxygen abundances of 12+log(O/H) = $8.02\pm 0.25$, $7.70\pm 0.06$, and $7.70\pm 0.10$ for Gz9p3, EGSY8p7, and ZD2, respectively, based electron temperatures $T_e \sim 1.5-1.9\times 10^{4}$\,K as summarised in Table~\ref{tab:properties}. Compared to their stellar masses and star-formation rates (SFRs), the galaxies are consistent in terms of their mass-metallicity and fundamental-metallicity relations to the typical high-redshift population \citep{Heintz23,Pollock26_FMR,Koller26}.

Although two of the main galaxies, Gz9p3 and ZD2, appear as typical bright star-forming galaxies, the exact nature of EGSY8p7 has been debated. For instance, it has been reported that the object is a supermassive black hole (SMBH) due to broad nebular H$\beta$ emission-line detected in the spectrum \citep{Larson23}. However, there is also evidence for broad [\oiii] lines from NIRSpec IFU data \citep{Zamora25}, with kinematics implying broadening due to ionised outflows driven by star-formation instead of an AGN. The object is also located in the star-forming locus of several diagnostic diagrams, further suggesting a lack of dominant AGN emission. The nitrogen lines present in the G140M spectrum  (\nv\,$\lambda1240$, \niv]\,$\lambda1486$, and \niii]\,$\lambda1747$) are broad and redshifted relative to other emission or absorption lines, which seems to confirm the ionised outflow scenario. More recently, it has also been argued that the \heii feature and the P-Cygni profiles observed for the high-ionisation lines \nv\ and \civ\ can be explained by very massive stars (VMS), with masses $100-300\, M_\odot$ \citep{MarquesChaves26}. Further, by modelling the two-photon and free-bound emission with {\tt PyNeb} \citep{Luridiana15} we here exclude a substantial contribution in EGSY8p7 from nebular continuum emission in the rest-frame UV, since the spectral slope and UV turnover are inconsistent with this particular model. This could otherwise bias the derived \hi\ column density \citep{Cameron24,Katz25,Terp24,Pollock26_DLA,Reumert26}. 

We emphasise that while the presence of VMS or an AGN in any of the three main galaxies may contribute to emission line measurements and therefore affect the emission line metallicities, we are agnostic to the underlying background emission for the absorption-derived values.

\subsection{Modelling the \lya\ damping wing} \label{sec:lya}

\begin{figure}
    \centering
    \includegraphics[width=0.85\linewidth]{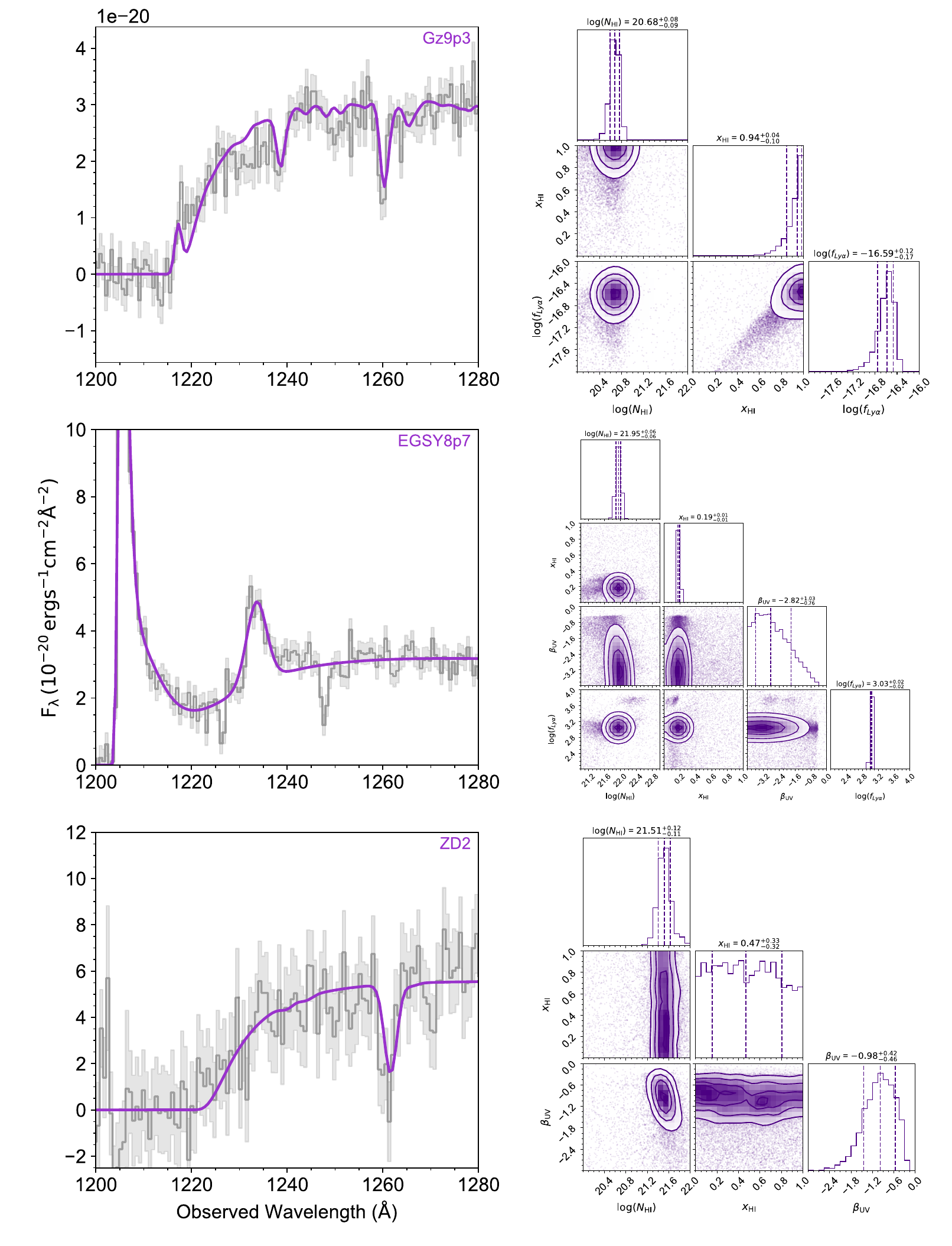}
    \caption{{\bf Best-fit model of the \lya\ damping wing profiles.} The JWST/NIRSpec rest-frame UV spectra for the three main galaxies are shown in gray, with best fit model for the DLA and IGM fitting overplotted in purple. The right panels show the corner plots for the posterior distributions for the main parameters: column density $N_{\mathrm{HI}}$, neutral IGM fraction $x_{\mathrm{HI}}$, UV spectral slope $\beta_{\mathrm{UV}}$, and \lya\ flux $f_{Lya}$. The derived column density $N_{\mathrm{HI}}$ for Gz9p3 is highly dependent on the \lya\ emission, with the model here showing the lower range of $N_{\mathrm{HI}}$ produced. See text for further details.
    }
    \label{fig:DLAfitting}
\end{figure}

We model the \lya\ turnover in the three main galaxy spectra with a DLA and IGM absorption, the former representing the contribution of local \hi\ gas in the ISM or CGM to the observed damping wing. We assume that the intrinsic rest-frame UV continuum can be expressed as a simple power-law function, $F_\lambda \propto \lambda^{-\beta}$, following the procedure described in Refs. \cite{Heintz24_DLA,Pollock26_DLA}. Using the stellar continuum model derived from \texttt{BAGPIPES} (see Sect.~\ref{sec:physprop}, and top panel of Figure~\ref{fig:DLAfitting}) instead has a limited impact on the derived \hi\ column densities ($\Delta N_{\rm HI}\lesssim 0.1$\,dex) and is less flexible in modelling the neutral hydrogen fraction in the IGM which is set to the expected redshift average by default. We also include components from the longer wavelengths absorption or emission lines in the model as Gaussian-line profile to more accurately fit the entire rest-frame UV continua. 

The DLA component is modelled by approximating the absorption-line profile as a Voigt function, following the method of Ref.~\citep{TepperGarcia06}. From this, a neutral hydrogen column density $N_{\mathrm{HI}}$ can be derived, which is mostly sensitive to the strength of the damping wing at high column densities. We also include a component for absorption from \hi\ in the IGM due to the Gunn-Peterson effect, using the formalism described in Refs. \citep{MiraldaEscude98, Totani06}. We set the lower redshift limit for the IGM absorption to be at $z=5.3$, when the large-scale reionisation is expected to be complete \citep{Bosman22}. 
The best-fit models for each galaxy are shown in Figure~\ref{fig:DLAfitting}, along with the corner plots showing the posterior distributions for the main parameters: column density $N_{\mathrm{HI}}$, neutral IGM fraction $x_{\mathrm{HI}}$, UV slope $\beta_{\mathrm{UV}}$, and \lya\ flux $f_{Lya}$.

There is evidence of weak \lya\ emission in Gz9p3, which here is modelled as a simple Gaussian. The column density derived is dependent on the inclusion of \lya\ emission and its width allowed in the modelling (a best-fit model including narrow \lya\ is shown in Figure~\ref{fig:CO_SiO}), ranging from $\mathrm{log}(N_{\mathrm{HI}}/\mathrm{cm^{-2}})\sim20.60 - 21.30$. We therefore report a conservative column density, of $\mathrm{log}(N_{\mathrm{HI}}/\mathrm{cm^{-2}})\sim21^{+0.3}_{-0.4}$, which represents the range of these models. We note this is at odds with the lower value derived in \citet{Chen26} with their ionised bubble model, finding $\mathrm{log}(N_{\mathrm{HI}})\sim19.5\,\mathrm{cm^{-2}}$.
When excluding \lya\ emission from the fit entirely, we also obtain a low column density, although it is unconstrained and dependent on the lower limit chosen for the prior. We note that such low \hi\ column densities would imply that the lower bound on the absorption-based metallicity for Gz9p3 is approximately solar, which we argue is highly unlikely and inconsistent with the emission-line measurement. Instead, we test a DLA-only model representing the maximum \hi\ column density in line the of sight, finding a similar column density to our adopted value of $N_{\rm HI} \sim 10^{21}\,\mathrm{cm^{-2}}$. 

For EGSY8p7, the \lya\ emission is extremely strong, showing a skewed spectral line profile, typically observed in high-redshift \lya\ emitters as an indication of significant resonance scattering \citep{Matthee20}. We model this line with both a Gaussian profile and an exponential (redward) wing, which accurately reproduces the observed line shape. Despite having broad and prominent \lya emission, EGSY8p7 still requires an extreme column density $N_{\mathrm{HI}}=21.95 \, \mathrm{cm^{-2}}$ in excess of the \hi\ contribution from the IGM to reproduce the lack of flux seen at $\lambda\sim 1220\mathrm{\AA}$ and the shape of the overall \lya\ turnover. 
Since there are strong nebular emission lines in EGSY8p7, it is possible that nebular continuum may contribute significantly to the spectrum. As two-photon emission at the UV turnover could be misinterpreted as a DLA, we test a nebular model with \texttt{PyNeb}. This can reproduce a moderate Balmer jump seen in the PRISM spectrum (with $T_e\sim30,000$\,K) and the UV slope above $\lambda\sim2000\mathrm{\AA}$, though the turnover is substantially lower than the observed flux, suggesting nebular continuum emission is not dominating the rest-UV spectrum, and cannot be significantly impacting the inferred column density. 

For ZD2, we derive a high column density ($\mathrm{log}(N_{\mathrm{HI}})=21.5\,\mathrm{cm^{-2}}$, also consistent with the value derived in Ref.~\citep{KeerthiVasan26}), with no apparent \lya\ emission and an unconstrained IGM fraction, likely due to the dominating DLA component. The derived \hi\ column densities are summarised in Table~\ref{tab:properties}. 

\subsection{Absorption-line measurements}

We model the absorption-line profiles observed in the three main galaxies and the DLA stack (see Sect.~\ref{sec:stack} below) with Voigt profiles, the convolution of a central Gaussian component and broad Lorentzian wings, using the software package {\tt VoigtFit} \citep{Krogager18}. This takes as input the observed spectra and the delivered spectral resolution ($\mathcal{R}\sim 1000$ for the JWST/NIRSpec G140M grating spectra) and outputs the column density $N$ and broadening parameter $b$ for each ionic transition. We model the most prominent absorption lines seen in the rest-frame UV spectrum; \sii\,$\lambda1260$, \oi\,$\lambda1302$, \oi*\,$\lambda 1304$, \sii\,$\lambda1304$, \cii\,$\lambda1334$, \cii*\,$\lambda1335.7$, \sii\,$\lambda1526$, and \alii\,$\lambda 1670$. These low-ion transitions all probe the metals in the neutral gas-phase, since their ionisation potential are below or equal to that of neutral hydrogen (13.6 eV). This effectively assumes that, for instance, $N_{\rm C,tot} = N_{\rm CII}$ in the neutral gas such that ionisation corrections are negligible due to total self-shielding of \hi\ from ionising photons as is the case for the strongest DLA systems \citep{Wolfe05}.  

Due to the medium resolution of the spectra, we are not able to resolve any potential sub-components in the transition of each ion. Consequently, this implies that the observed optical depth of each transition will reflect the total integrated column density and the widths the total turbulence or line dispersion, quantified via the broadening `$b$'-parameter. Further, the instrumental broadening will also make most lines appear unsaturated or subject to low covering fractions, even though the intrinsic line profiles are actually mildly saturated. To break this degeneracy, we assume fixed $b$-parameters of $b=50$\,km\,s$^{-1}$ for Gz9p3, and $b=100$\,km\,s$^{-1}$ for EGSY8p7 and ZD2, respectively, to best reproduce the observed line profiles for each source and for more conservative abundance estimates. The derived column densities are listed in Table~\ref{tab:coldens}. We obtain consistent results from the measured equivalent widths (EWs) for each line transition, with $N = 1.13\times 10^{20} \, {\rm EW_{rest}} / (f_{\rm osc}\lambda^2)$, where $f_{\rm osc}$ is the oscillator strength and $\lambda$ the rest-frame wavelength of each line, respectively, in the optical thin regime. The derived column densities are generally insensitive to the exact $b$-parameter assumed (within $\Delta N<0.2$\,dex) for the range $b=20-100$\,km\,s$^{-1}$. We caution that for any intrinsically narrower components ($b<20$\,km\,s$^{-1}$), the output column densities will be substantially higher, due to a significant `hidden' intrinsic saturation in the lines. We argue, however, that such narrow components are unphysical in this extended, turbulent medium given also that the observed lines likely trace several velocity sub-components.   

\begin{table}[]
    \centering
    \begin{tabular}{c c c c c} 
       \hline\hline
        & $\log N$(C\,{\sc ii}) & $\log N$(O\,{\sc i}) & $\log N$(Si\,{\sc ii}) & $\log N$(Fe\,{\sc ii}) \\
        \hline
        Gz9p3 & $16.22\pm 0.16$ & $16.23\pm 0.24$ & $14.80\pm 0.13$ & $<14.62$ \\
        EGSY8p7 & $15.76\pm 0.11$ & $14.93\pm 0.13$ &  $13.81\pm 0.10$ & $<14.49$ \\
        ZD2 & $17.00\pm 1.01$ & $15.87\pm 0.45$ & $14.16\pm 0.81$ & $<14.73$ \\
        \hline
    \end{tabular}
    \caption{The derived column densities for the three main galaxies from {\tt VoigtFit}. The values are in units of cm$^{-2}$, and the \feii\ column densities are reported as $3\sigma$ upper limits.}
    \label{tab:coldens}
\end{table}

In all cases, we are able to resolve the main ground-state components from the first excited states, for example \cii\,$\lambda1334$ from \cii*\,$\lambda1335.7$ and \oi\,$\lambda1302$ from \oi*\,$\lambda 1304$. Crucially, we find that not including the excited states in the modelling provides a statistically significant worse fit to the data. While the \cii\,$\lambda1334$ and \cii*\,$\lambda1335.7$ line complex is uncontaminated by other line transitions at the same redshift, the inclusion of \oi*\,$\lambda 1304$ will impact the modelling of the \oi\,$\lambda1302$ and \sii\,$\lambda1304$ line profiles. The latter transition has been assumed to trace the covering fraction when compared to \sii\,$\lambda1260$ for Gz9p3 \citep{Chen26}, though this will depend heavily on the relative contribution of \oi*\,$\lambda 1304$ to the optical depth, which is not possible to resolve from \sii\,$\lambda1304$ at the delivered resolution. The added constraints from the \sii\,$\lambda1260$ and \sii\,$\lambda1526$ line profiles for the derived $N_{\rm SiII}$ help break this degeneracy and allow for a more accurate measurement of the column density of atoms in the \oi*\,$\lambda 1304$ state. This yields the total O abundance when added to the derived ground-state \oi\,$\lambda1302$ column density.

Since the measured absorption lines column densities are insensitive to the exact ionisation and density state, we are agnostic to the underlying background continuum source in our modelling. For instance, EGSY8p7 has been argued to either host an active galactic nuclei (AGN; \citep{Larson23}) or have a stellar population dominated by very massive stars up to $300\,M_\odot$ \citep{MarquesChaves26}. Further, while we reproduce the slight blueshift of the absorption-line redshift relative to that derived from the emission lines for Gz9p3 \citep{Chen26}, we argue that this is also consistent with the proposed scenario of the neutral gas absorption tracing more pristine regions in the kinematically distinct outskirts of the galaxy (potentially in the CGM; \citep{Zhu26}). 

While the above assumptions and caveats imply that the derived gas-phase metallicities should potentially be treated as lower limits, we argue that the measured [C/O] abundance ratios are robust. First of all, high-resolution GRB and quasar absorption-line spectroscopy show that the velocity components of \cii\,$\lambda1334$ are identical to the ones seen for \oi\,$\lambda1302$ in the line of sight through foreground galaxies at $z=2-6$ \citep{Heintz21,Wilson24}. This suggests that all the \cii\,$\lambda1334$ and \oi\,$\lambda1302$ velocity components and sub-structures co-exist within the same neutral gas.  The redshift evolution of the \cii\ and \oi\ line densities have also been found to be consistent at $z>5$ with \jwst\ quasar absorption-line spectroscopy \citep{Christensen23}. This implies that the covering fraction must be near-identical. Physically, most of the carbon in the neutral gas will be in the form of \cii\ due to the low ionisation potential of \ci\ of 11.2 eV, co-existing with \oi\ (with ionisation potential of 13.62 eV) and shielded from more energetic ionising photons by the dense \hi\ gas in the photo-dissociation region \citep{Hollenbach99}. This is also seen in their consistent absorption-line profiles, whereas the higher ionisation lines \siv\ and \civ\ appear slightly blueshifted due to presence of prominent P-Cygni profiles \citep{MarquesChaves26}. Their depletion strengths onto dust are also very similar \citep{Konstantopoulou24}, minimising any potential depletion biases in their relative abundance measurements. Finally, they also have similar oscillator strengths and solar abundances, such that any potential hidden saturation will affect the lines almost equally. The relative equivalent widths of optical depths of the \cii\ and \oi\ absorption lines are thus a remarkable tracer of the actual intrinsic line ratio of the two, making the derived relative [C/O] abundances robust to most model assumptions and caveats highlighted above.  

While the individual galaxies may appear to be carbon enhanced relative to oxygen, the CEMP region in stellar archaeology is defined as [C/Fe] $>1.0$ \citep{Beers05}. The absence of clear Fe absorption features means we can only obtain lower limits based on the non-detection of \feii\,$\lambda1608$. For Gz9p3, EGSY8p7, and ZD2 we obtain [C/Fe] values $>0.59$, $>0.26$, and $>0.32$ respectively at $3\sigma$ significance ($>0.99$, $>1.29$, $>1.19$ at $1\sigma$ significance). 
There is no strong evidence that these sources are indeed significantly carbon-enhanced relative to iron, though it is difficult to confirm with the current data available, and lack of detection of \feii\,$\lambda1608$. 
We also note that Fe and Si are refractory elements and are thus more prone to dust depletion than the more volatile C or O \citep{JenkinsWallerstein17, RomanDuval22}, so both the Fe and Si abundances could be underestimated if depleted from the gas-phase onto dust grains. 


\subsection{Spectral stacking} \label{sec:stack}

To complement the spectra for the three main individual sources observed as part of the SPURS programme, we compile all the publicly available \jwst/NIRSpec G140M spectra on DJA of galaxies at $z>8$ to test if absorption features can be detected in an average stack. There are a total of 32 unique spectra, from a range of programmes including JADES (PIDs: 1181, 1286, 1287, 3215, 1210), SPURS (9214), GO-4750, and GO-5943. We de-redshift each spectrum and resample the flux along a common wavelength grid, which is set by the resolution of the lowest-redshift galaxy and its delivered spectral resolving power. A weighted average is then used to stack the resampled spectra.

\begin{figure}
    \centering
    \includegraphics[width=1.0\linewidth]{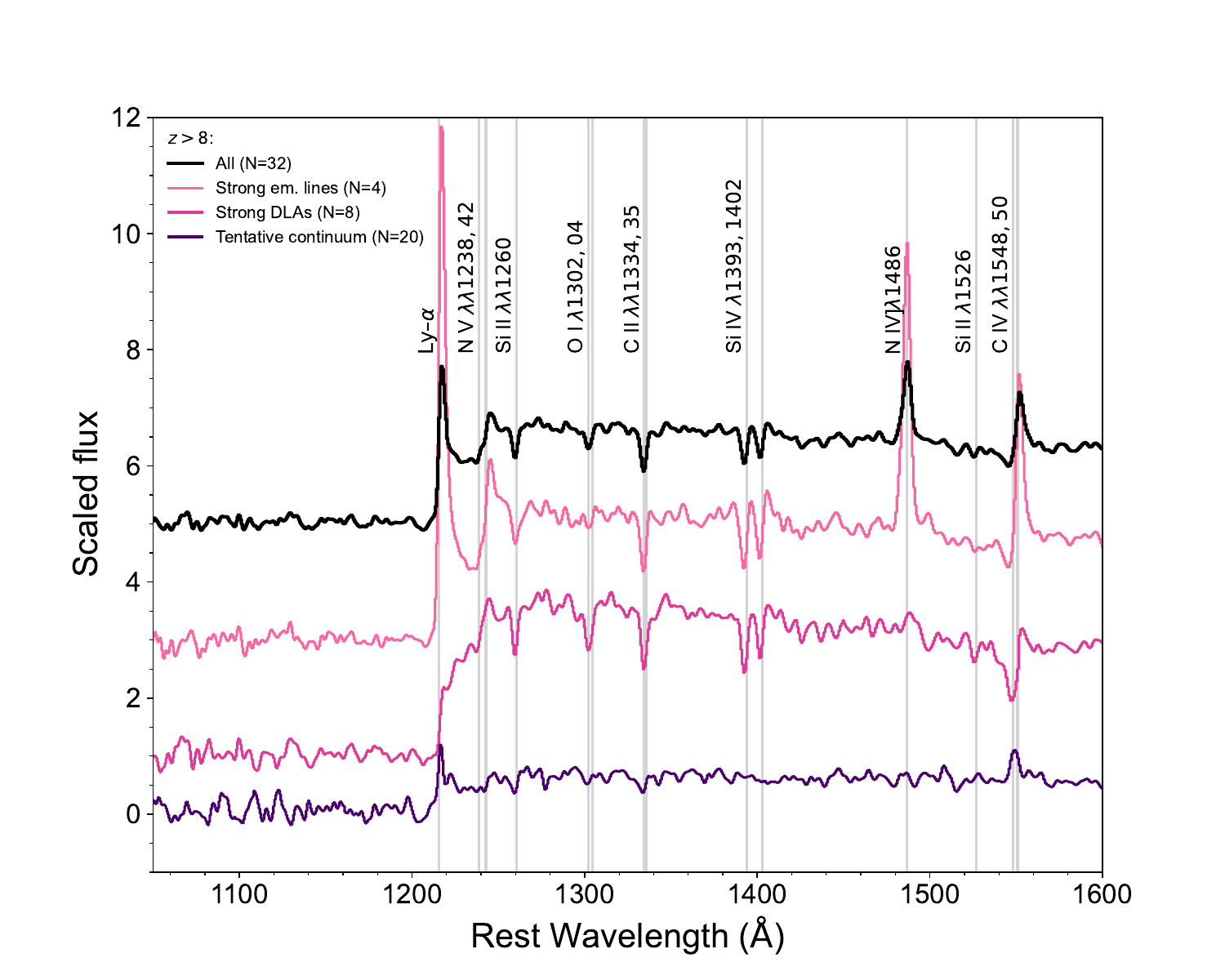}
    \caption{{\bf Weighted stacks of publicly available $z>8$ JWST/NIRSpec G140M spectra}. These stacks include data from JADES (PIDs: 1181, 1286, 1287, 3215, 1210), SPURS (9214), GO-4750, and GO-5943. The total stack is shown in black, with additional stacks for strong line emitters ($N=4$), strong DLAs/good continuum detections ($N=8$), and tentative or no clear features in the continuum ($N=20$). The location of potential low-ion absorption lines and selected emission lines are highlighted.}
    \label{fig:stacked}
\end{figure}

The average stack of the 32 included sources is shown in Figure~\ref{fig:stacked}, with \lya\ emission, clear absorption features for \sii\,$\,\lambda1260$, \oi\,$\,\lambda1302$, \cii\,$\,\lambda1334$, and \siv$\,\lambda\lambda1393,1402$, emission in nitrogen \niv]\,$\,\lambda1486$, and a potential P-Cygni feature for \civ\,$\,\lambda\lambda1548,50$.  We also group the spectra into strong line emitters ($N=4$), strong DLAs or good continuum detection ($N=8$), and those with only tentative continuum detection ($N=20$). The strong line emitter sample also exhibits strong \lya\ emission, and a clear P-Cygni profile in \civ. The DLA sample shows a prominent \lya\ turnover and strong absorption lines, with a lack of prominent emission lines. The strong absorption features are expected alongside a high covering fraction of \hi\ gas as the low-ion metal absorption lines trace the neutral gas phase. The final stack shows a lack of any strong features, with only tentative detection of \lya\ and \civ\ emission. The majority of the galaxies used in this stack have short exposure times and lack the necessary depth to detect absorption components. 

We derive an average \hi\ column density for the DLA stack, while for the total stack we are unable to obtain a robust posterior on $N_{\rm HI}$ due to contamination from \lya\ emission. We also calculate the average [C/O] and [Si/O] abundance ratios for the DLA and total stacks. 
Since the stacked absorption features will be a blend of several independent line profiles and velocity components we are unable to use {\tt VoigtFit} to model the absorption line feature, so we instead measure rest-frame equivalent widths (EWs) and convert to column densities assuming an optically thin regime, and a directly proportional relationship\citep[see also][]{Zhu26}. We calculate the expected contribution of \sii\,$\,\lambda1304$ using the derived column density of \sii\,$\,\lambda1260$, and subtract this from the measured EW around the \oi$\,\lambda1302$ line profile. The resulting stacked ratios of [C/O] are shown in Figure~\ref{fig:CO_OH} for the total and DLA stacks, with values of $\mathrm{[C/O] = 0.02\pm0.20}$ and $-0.10 \pm 0.13$ respectively, both exhibiting close to solar carbon to oxygen ratios, though lower than the three individual measurements. The stacked ratios of [Si/O] obtained are also slightly lower than the main galaxies, with [Si/O] $=-0.24\pm0.15$ and $-0.31\pm0.14$ for the total and DLA stacks respectively. 


\subsection{Insights into the nature of Pop\,III SNe from simulations}


To interpret the gas-phase chemical abundances derived here and the stellar progenitors they are synthesised from, we consider the predictions from the tailored {\tt NEFERTITI} simulations \citep{Koutsouridou26} and the parametric model from Ref. \citep{Vanni23}. Briefly, {\tt NEFERTITI} is a semi-analytical galaxy-evolution model aimed at studying the first stars and their distinct chemical signatures, making it ideally suitable to interpret our observations. {\tt NEFERTITI} is implemented onto the {\it Caterpillar} suite \citep{Griffen16} of dark matter only cosmological simulations of Milky Way-like galaxies. By specifically constraining the model to reproduce local Galactic data, 
it provides robust, locally calibrated predictions for the chemical enrichment initiated by the first stars. Specifically, the galaxy-formation model follows individual Pop\,III stars forming in pristine gas, and self-consistently tracks the chemical composition of the gas according to the individual Pop\,III supernovae (SNe), up to and including the stellar yields from more standard Pop\,II supernovae. 
The assumed initial mass function (IMF) for the Pop\,III stars is a Larson-type with stellar masses in the range $[0.8-1000] \, M_\odot$, with a characteristic mass of $10 \, M_\odot$. This is compatible with both local \citep{Koutsouridou24} and high-redshift constraints \citep{Rusta26}. Pop\,III stars in the mass range $[10-100] \, M_\odot$ can explode as SNe with different energies, following an energy distribution function calibrated to reproduce the fraction of CEMP stars in the Milky Way halo, whereas Pop\,III stars with initial masses $[140, 260] \, M_\odot$ explode as pair-instability SNe (PISNe). 
Here, we adopt the same notation of \citep{Koutsouridou26}: Faint SNe ($E_{\rm SN} = 0.3 - 0.6 \times 10^{51}\, {\rm erg}$), CCSNe ($E_{\rm SN} = 0.9 - 1.5 \times 10^{51}\, {\rm erg}$), high energy SNe ($E_{\rm SN} = 1.8 - 3 \times 10^{51}\, {\rm erg}$), and hypernovae ($E_{\rm SN} = 5 - 10 \times 10^{51}\, {\rm erg}$). 
Throughout, {\tt NEFERTITI} assumes the Pop\,III SNe yields from Ref. \citep{Heger10} and PISNe yields from Ref. \citep{Heger02}.

To most directly compare with the absorption-line properties observed here, we consider the chemical yields in the gas-phase of the simulated galaxies in {\tt NEFERTITI}. These are physically different from the predictions shown in \citep{Rusta25}, which are emission lines obtained from synthetic galaxy spectra.
For the contours plotted in Figure~\ref{fig:CO_OH}, we only consider star-forming galaxies at redshifts $7.5<z<9.5$ to more accurately represent the three main sources. The simulated data is further split into separate populations, for galaxies that are either predominantly enriched by Pop\,III or Pop\,II supernovae at different mass or energy ranges, where the major contributions are coming from either faint, high-energy, or pair-instability supernovae. The 90\% contours and the scatter points representing each of the simulated galaxies show that the faint Pop\,III-enriched galaxies are the only ones that reproduce the increasing C/O abundances at low metallicities. The evolutionary tracks for the C/O abundance ratio as a function of metallicity from Kobayashi \& Ferrara \citep{Kobayashi24} shown in Figure~\ref{fig:CO_OH} also predict a similar upturn at metallicities 12+log(O/H) $=6.2-7.0$, depending on the main underlying stellar population considering for instance a top-heavy Pop\,II IMF with Wolf-Rayet stars or a Pop\,III stellar IMF. We caution that this model use different yields from Ref.~\citep{Kobayashi20}, though we find that the results are consistent with the {\tt NEFERTITI} models.

\begin{figure}
    \centering
    \includegraphics[width=1.0\linewidth]{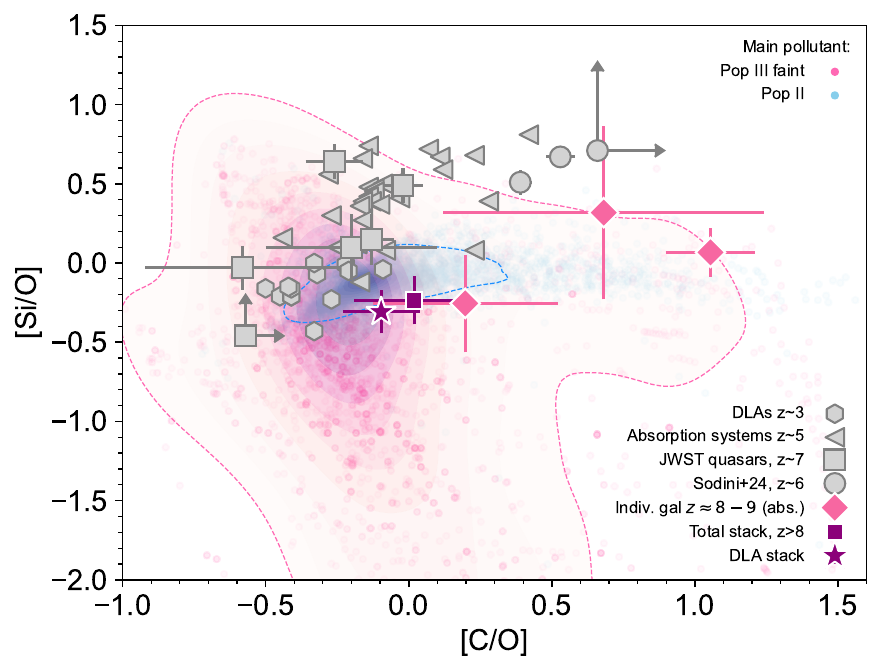}
    \caption{\textbf{Silicon and carbon abundances relative to oxygen.} The symbol notation follows Fig.~\ref{fig:CO_OH}. The absorption-line systems for the three main galaxies show high [C/O] but generally low silicon [Si/O] abundance ratios, placing them in the regime where the {\tt NEFERTITI} simulations predict either Pop\,II supernovae or Pop\,III faint supernovae to dominate the chemical yields. The average values derived from the total and DLA stacks of G140M spectra at $z>8$ are also shown. We compare to literature values of DLAs at $z\sim3$ \citep{Cooke11, Cooke17, Welsh19}, absorption-line systems at $z\sim5$ \citep{Becker19, Cooper19}, and quasars at $z\sim7$ \citep{Christensen23, Vanni24, Sodini24}, which generally show lower carbon abundance at fixed Si/O.}
    \label{fig:CO_SiO}
\end{figure}

Based on the [C/O] versus [O/H] contours of the predicted yields from the {\tt NEFERTITI} simulations, we find that two of the three main galaxies, Gz9p3 and ZD2, are consistent with both the faint Pop\,III- and the Pop\,II-enriched phase, whereas EGSY8p7 is occupying the region with yields predicted solely from the faint Pop\,III CCSNe. To further investigate the likely progenitor stars driving the observed chemical enrichment we consider the Si/O abundance ratios relative to C/O specifically for the faint Pop\,III- and the Pop\,II-enriched systems in Fig.~\ref{fig:CO_SiO}. This is based on the formalism described in Ref. \citep{Vanni24} and reproduced with the updated {\tt NEFERTITI} models shown here.
Although the three main galaxies have high C/O, the derived Si/O ratios are lower, also below those recently reported in Ref.~\cite{Nakane26}, though the [C/O] values are comparable. 
While we are able to exclude a scenario with yields from PISNe with typical high Si/O but low C/O abundance ratios, all three of the main galaxies are consistent with the expected elemental abundances from both the Pop\,II supernovae or Pop\,III faint supernovae. Searching the full simulated sample for matches in the [O/H], [C/O], and [Si/O] abundance patterns supports the ambiguity in the solution for Gz9p3 and ZD2. However, the source EGSY8p7 can only be matched to a system with a faint Pop\,III CCSNe enrichment, driven mainly by its low metallicity.
Both in Fig. \ref{fig:CO_OH} and Fig. \ref{fig:CO_SiO}, the system lies at the edges of the Pop\,III model predictions. Indeed, {\tt NEFERTITI} produces predictions for entire galaxies, which are usually chemically enriched by tens of supernovae. The EGSY8p7 system is consistent with rarer models, where only a few faint SNe explode in the galaxy. Indeed, the observed features could be coming from more pristine regions of gas rather than representing whole galaxies, which is also supported by the deviations for the carbon and oxygen abundances derived for the central star-forming region in emission. This faint Pop\,III supernovae enrichment for the three main galaxies is further supported by the models of Ref. \citep{Vanni24}, which produce the chemical composition expected from individual stellar progenitors. The predictions of this simpler parametric model cover a wider range of the [Si/O] vs [C/O] space, with the three main galaxies occupying only the faint Pop\,III regime, highlighting this scenario as the likely stellar progenitors of the elemental abundances observed in absorption in this work. 

\end{document}